\documentclass[11pt,bm,aps,showpacs,nofootinbib,amsfonts,amssymb,preprintnumbers]{revtex4}

 \usepackage{here}

\usepackage[dvipdfmx]{graphicx}
\usepackage{amssymb,amsfonts,amsmath,bm,color,multirow,cases,empheq,hyperref}
\usepackage{mathrsfs}

\newcommand{\blue}[1]{{\color{blue} #1}}
\usepackage{enumerate}
\usepackage{ulem}
\usepackage{afterpage}

\hypersetup{%
 setpagesize=false,
 bookmarksnumbered=true,%
 bookmarksopen=true,%
 colorlinks=true,%
 linkcolor=blue,%
 citecolor=red}

\begin{document}




\title{A black lens in bubble of nothing
}

\author{Shinya Tomizawa}
\email{tomizawa@toyota-ti.ac.jp}
\author{Ryotaku Suzuki}
\email{sryotaku@toyota-ti.ac.jp}
\affiliation{Mathematical Physics Laboratory, Toyota Technological Institute\\
Hisakata 2-12-1, Nagoya 468-8511, Japan}
\date{\today}

\preprint{TTI-MATHPHYS-15}




\begin{abstract} 
Applying the inverse scattering method to  static and bi-axisymmetric Einstein equations, we construct a non-rotating black lens inside bubble of nothing whose horizon is topologically lens space $L(n,1)=S^3/{\mathbb Z}_n$. 
Using this solution, we discuss whether a static black lens can be in equilibrium by the force balance between the expansion and gravitational attraction. 

  \end{abstract}

\date{\today}
\maketitle



\section{Introduction}

The studies on higher-dimensional black hole solutions to Einstein's equations have played important roles in the microscopic derivation of black hole entropy~\cite{Strominger:1996sh} and in the fundamental research on the scenario of large extra dimensions~\cite{Argyres:1998qn} through  black hole production in an accelerator.  
The recent developments in solution-generation techniques enable us to find various exact solutions of higher-dimensional black holes, our understanding of them is still not enough. 
For example,  according to the topology theorem for five-dimensional black holes~\cite{Galloway:2005mf,Cai:2001su,Hollands:2007aj,Hollands:2010qy},  the topology of the spatial cross section of the event horizon must be either sphere $S^3$, ring $S^1\times S^2$ or lens spaces $L(p,q)$, if the spacetime with  commuting two rotational Killing vector fields and a timelike Killing vector field
 is asymptotically flat.  
 As for the first two topologies, the exact solutions to vacuum Einstein's equations~\cite{Tangherlini:1963bw,Myers:1986un,Emparan:2001wn,Pomeransky:2006bd} have already been found. 
For lens space topology, however, it has been difficult to find a regular vacuum solution since the resultant solutions have naked singularities.

\medskip
The inverse scattering method (ISM) is perhaps one of the most powerful tools to obtain exact solutions of Einstein equations with $(D-2)$ Killing isometries, where $D$ is a spacetime dimension. 
In particular, combined with the concept of the rod structure~\cite{Harmark:2004rm}, this method has succeeded to derive five-dimensional vacuum black hole solutions. 
The first example of the construction of black hole solutions by the ISM is the re-derivation of the five-dimensional Myers-Perry black hole solution~\cite{Pomeransky:2005sj}.  
Next, the  black ring with $S^2$-rotation  was  re-derived~\cite{Tomizawa:2005wv} by the ISM (this was first derived in Refs.~\cite{Mishima:2005id,Figueras:2005zp}), 
but it turned out that the generation of the black ring with $S^1$ rotation  has a certain problem on how to choose the seed, since an easy choice of the seed always results in the generation of a singular solution. 
The suitable seed to derive the black ring with $S^1$-rotation was first considered in~\cite{Iguchi:2006rd,Tomizawa:2006vp}. 
Subsequently, the more general black ring solution with both $S^1$ and $S^2$ rotations was constructed by Pomerasnky and Sen'kov~\cite{Pomeransky:2006bd}.

\medskip
Using the ISM, some authors attempted to construct asymptotically flat black lens solutions to the five-dimensional vacuum Einstein equations. 
First, Evslin~\cite{Evslin:2008gx} attempted to construct a static black lens with the lens space topology of $L(n^2+1,1)$ but  found that curvature singularities cannot be eliminated, whereas both conical and orbifold singularities can be removed.  
Moreover,  by using the ISM, Chen and Teo~\cite{Chen:2008fa} constructed  black hole solutions with the horizon topology of $L(n,1)=S^3/{\mathbb Z_n}$ but these solutions must have either conical singularities or naked curvature singularities. 
Thus, the major obstacle in constructing a black lens solution is always the existence of naked singularities. 
However, the sudden breakthrough  has come from supersymmetric solutions. 
Kunduri and Lucietti~\cite{Kunduri:2014kja} succeeded in the derivation of the first regular supersymmetric solution of an asymptotically flat black lens with the horizon topology of $L(2,1)=S^3/{\mathbb Z}_2$. 
This solution was further generalized to the more general supersymmetric black lens with the horizon topology  $L(n,1)=S^2/{\mathbb Z}_n\ (n\ge 3)$~\cite{Tomizawa:2016kjh,Breunholder:2017ubu}.  
Getting a useful clue from the work of Kundhuri and Lucietti, Ref~\cite{Tomizawa:2019acu} attempted to construct the vacuum solution of the black lens with $L(2,1)$ without singularities but the solution has unavoidable closed timelike curves (CTCs). 
Thereafter, Ref.~\cite{Lucietti:2020phh} discuss nonexistence of vacuum black lenses.

\medskip

Supersymmetric black lenses carry the mass, electric charge (saturating BPS bound), two angular momenta and magnetic fluxes~\cite{Kunduri:2014kja,Tomizawa:2006vp}.
As is discussed in Ref.~\cite{Tomizawa:2006vp}, there exists no limit such that all the magnetic fluxes vanish. 
Therefore, as for the supersymmetric solutions, the existence of the magnetic fluxes seem to play an essential role in supporting the horizon of the black lens. 
In general, however, it is not clear whether such magnetic fluxes necessarily need to construct a black lens.
Recently,  a different type of solutions within a class of generalized Weyl solutions, static black hole binaries and black rings in expanding bubbles of nothing, was studied in Ref.~\cite{Astorino:2022fge}, although so far equilibrium configuration of black holes in bubble  had been  studied  in the context of  Kaluza-Klein theory~\cite{Elvang:2002br,Tomizawa:2007mz,Iguchi:2007xs}. 
As is well-known, an asymptotically flat, static black ring cannot be in equilibrium since the horizon collapse  due to the self-gravitational force. 
However, the black ring in~\cite{Astorino:2022fge} is allowed to be in static  equilibrium by the balance between the expanding force of a bubble and the gravitational force, so it has no conical singularities. 
This solution leads a simple,  interesting question to us.
Is a non-rotating black lens in bubble of nothing allowed to be  in  equilibrium? 
To study such a solution may enable us to know  what (except for magnetic fluxes) is needed to obtain a regular black lens.
Thus, the end of this paper is to investigate whether expanding bubble of nothing admits the existence of a black lens in equilibrium. 
In this paper, to derive such a solution, we apply the ISM to the five-dimensional vacuum Einstein equations with staticity and  bi-axisymmetry, and construct a $1$-soliton solution by considering a static black ring inside of bubble in~\cite{Astorino:2022fge} as a seed solution. 
Note that our procedure in the ISM is entirely the same as the work of Chen-Teo where the seed solution is chosen as a static black ring, namely, the only different point is the seed solution.

\medskip
We organize the remaining part of this paper as follows. 
In the following section~\ref{sec:solution}, under the assumptions of  staticity and bi-axisymmetry,  we present a vacuum solution of a non-rotating black lens with the horizon topology $L(n,1)$ in bubble of nothing  as a $1$-soliton solution in five dimensions by using the ISM. 
In section~\ref{sec:boundary}, we impose  the boundary conditions under such that the spacetime has
 none of curvature, conical, and orbifold singularities on the axis and horizon.
 In section~\ref{sec:ctcs}, we further impose no closed timelike curves (CTCs) in the domain of communication. 
 In section~\ref{sec:black lens}, we discuss whether the non-rotating black lens in bubble of nothing indeed exist. 
  In section~\ref{sec:limit}, we confirm the limit of our solution to the Chen-Teo static solution.
  In the final section~\ref{sec:summary}, we devote ourselves to the summary and discussion on our results.



\section{ Static black ring in bubble of nothing as a seed solution}\label{sec:solution}
In general, the metric for a stationary and bi-axisymmetric spacetime can be written in the canonical coordinates as
\begin{eqnarray}
ds^2=g_{ab}dx^adx^b+f(d\rho^2+dz^2),\quad (a,b=t,\phi,\psi),
\end{eqnarray}
where $g_{ab}$ and $f$ depend on only $(\rho,z)$. 
The following constraint condition must be satisfied 
\begin{eqnarray}
{\rm det} (g_{ab})=-\rho^2. \label{eq:det}
\end{eqnarray}

\begin{figure}[H]
 \centering
\includegraphics[width=10cm]{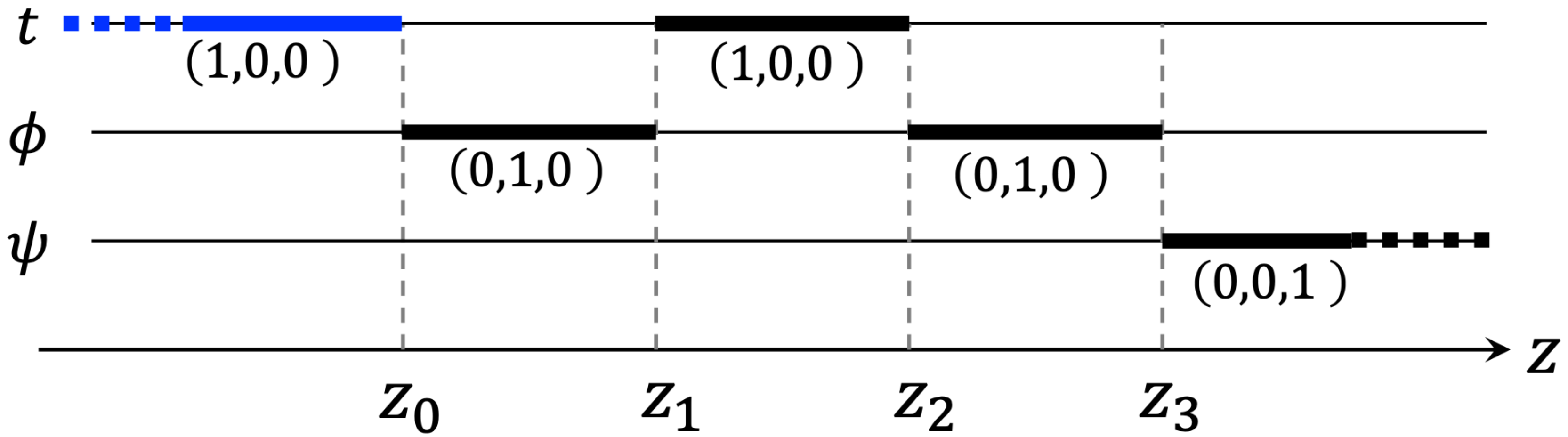}
\caption{Rod structure of the black ring inside bubble of nothing. }
\label{fig:rod}
\end{figure}

According to the procedure of Chen and Teo~\cite{Chen:2008fa}, we construct the static black lens in bubble of nothing by the ISM, where the static black ring as the seed solution is replaced with the black ring in bubble of nothing~\cite{Astorino:2022fge}\ (see Fig.~\ref{fig:rod} on the rod structure.). 

\medskip
Therefore, let us start with the exact solution of the black ring in bubble of nothing to the five-dimensional vacuum Einstein equations, whose metric is given by
\begin{eqnarray}
G_0&=&{\rm diag\ }\left(-\rho^2 \frac{\mu_1}{\mu_0\mu_2},\frac{\mu_0\mu_2}{\mu_1\mu_3},\mu_3 \right),\\
f_0&=&C_f\frac{\mu_3W_{01}^2 W_{03} W_{12}^2 W_{23}}{W_{02}^2W_{13} W_{00} W_{11} W_{22} W_{33}},
\end{eqnarray}
where for $i,j=0,1,2,3$, 
\begin{eqnarray}
\mu_i&:=&\sqrt{\rho^2+(z-z_i)^2}-(z-z_i), \\
\bar \mu_i&:=&-\frac{\rho^2}{\mu_i},\\
W_{ij} &:=&\rho^2+\mu_i\mu_j.
\end{eqnarray}
First, let us remove a trivial anti-soliton with the BZ vector $(0,0,1)$ at $z=z_3$:
\begin{eqnarray}
g_0&=&{\rm diag\ }\left(1,1,-\frac{\bar \mu_3^2}{\rho^2}\right)  G_0={\rm diag\ }\left(1,1,-\frac{\rho^2}{\mu_3^2}\right)  G_0,
\end{eqnarray}
In turn, let us add back a non-trivial anti-soliton with $(0,-a,1)$ to this, and then we can obtain a new $1$-soliton solution, a solution of {\it a static black lens in bubble of nothing}, 
\begin{eqnarray}
g_1&=&-\rho^2 \frac{\mu_1}{\mu_0\mu_2}dt^2
+\frac{\mu_0 \mu_2 ( \mu_1W_{03}^2W_{23}^2 +  4a^2z_3^2\mu_0\mu_2\mu_3^2 W_{13}^2 )}
{\mu_1\mu_3  ( \mu_1 W_{03}^2W_{23}^2 -  4a^2z_3^2\rho^2 \mu_0\mu_2 W_{13}^2 )}d\phi^2\notag\\
&&-2az_3\frac{ 2  \mu_0\mu_2  W_{03} W_{13} W_{23}W_{33}   }
{  \mu_3  ( \mu_1 W_{03}^2W_{23}^2 -  4a^2z_3^2\rho^2 \mu_0\mu_2 W_{13}^2 )  } d\phi d\psi \notag\\
&&+\frac{   \mu_1 \mu_3^2 W_{03}^2 W_{23}^2     +       4a^2 z_3^2\rho^4 \mu_0\mu_2 W_{13}^2   }
{ \mu_3  ( \mu_1 W_{03}^2W_{23}^2 -  4a^2z_3^2\rho^2 \mu_0\mu_2 W_{13}^2 )} d\psi^2, 
\end{eqnarray}
and 
\begin{eqnarray}
f_1
   &=& f_0 \frac{ \mu_1 W_{03}^2W_{23}^2 -  4a^2z_3^2\rho^2 \mu_0\mu_2 W_{13}^2  }
   {\mu_1 W_{03}^2 W_{23}^2 }.
\end{eqnarray}
It is easy to confirm that in the limit of $a\to0$, this solution  coincides with the static black ring in bubble of nothing~\cite{Astorino:2022fge}. 
One should note that $t,\psi$ is dimensionless and $\phi$ has the dimension of length. 
In the following section, after an appropriate coordinate transformation, we will impose the periodicity of $\phi, \psi$ so that conical singularities do not exist on symmetry of axises.

\section{Boundary conditions on the rods} \label{sec:boundary}
\begin{figure}[H]
 \centering
\includegraphics[width=10cm]{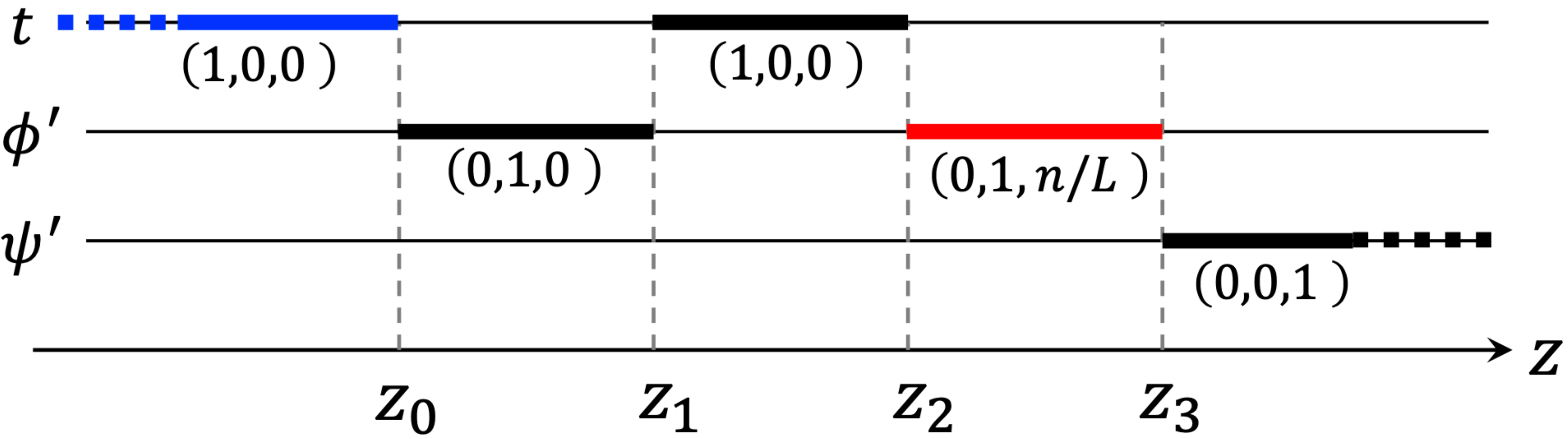}
\caption{Rod structure of the black lens inside bubble of nothing. }
\label{fig:rod-lens}
\end{figure}

In order to impose the appropriate boundary conditions so that the solution has the rod structure in Fig.~\ref{fig:rod-lens} and  neither conical singularities nor orbifold singularities,  
let us introduce new parameters $b:=z_3a$  and new coordinates $(\phi',\psi')$ defined by
\begin{eqnarray}
\frac{\partial}{\partial \phi'}=\sqrt{\frac{z_{30}}{z_{30}-2b^2}} \left(\frac{\partial}{\partial \phi}+\frac{b}{z_{30}}\frac{\partial}{\partial \psi}\right),\quad 
\frac{\partial}{\partial \psi'}=\sqrt{\frac{z_{30}}{z_{30}-2b^2}} \left(\frac{\partial}{\partial \psi}+2b\frac{\partial}{\partial \phi}\right),
\end{eqnarray}
and then the angular components are written as
\begin{eqnarray}
g_{\phi'\phi'}&=&\frac{z_{30}}{z_{30}-2b^2}\left(g_{\phi\phi}+\frac{b^2}{z_{30}^2} g_{\psi\psi}+\frac{2b}{z_{30}} g_{\phi\psi} \right), \\
g_{\psi'\psi'}&=&\frac{z_{30}}{z_{30}-2b^2}(g_{\psi\psi}+4b^2 g_{\phi\phi}+4b g_{\phi\psi}),\\
g_{\phi'\psi'}&=&\frac{z_{30}}{z_{30}-2b^2}\left[   \left( 1+\frac{2b^2}{z_{30}}\right) g_{\phi\psi}+2bg_{\phi\phi}+\frac{b}{z_{30}} g_{\psi\psi} \right],
\end{eqnarray}
where it should be noted that the constraint condition~(\ref{eq:det}) is preserved.

\medskip
Focusing on the two-dimensional space $\Sigma=\{(\rho,z)|\ \rho> 0,-\infty<z<\infty\}$, let us  study the rod structure of the obtained solution. 
The rod structure~\cite{Emparan:2001wk,Harmark:2004rm} enables us to understand stationary and axisymmetric solutions (more precisely,  solutions with $(D-2)$ commuting Killing vectors) easily in a diagrammatic may.
The $z$-axis ($\rho=0$) of  the metric, which corresponds to a fixed point set of a certain Killing isometry, is decomposed into five parts, $I_-=\{(\rho,z)\ | \ \rho=0,\ z<z_0\}$, $I_0=\{(\rho,z)\ | \ \rho=0,\ z_0<z<z_1\}$, $I_1=\{(\rho,z)\ | \ \rho=0,\ z_1<z<z_2\}$, $I_2=\{(\rho,z)\ | \ \rho=0,\ z_2<z<z_3\}$, $I_+=\{(\rho,z)\ | \ \rho=0,\ z_3<z\}$. 
Thus, the boundary $\partial \Sigma$ of $\Sigma$ are composed of $I_\pm,I_i\ (i=0,\ldots,2)$ and the asymptotic region $I_\infty=\{(\rho,z)|\ \sqrt{\rho^2+z^2}\to \infty \ $ with $ z/\sqrt{\rho^2+z^2}$ finite$\}$.

\medskip
Now, we impose the conditions on each rod so that the solution has the same rod structure as Fig.~\ref{fig:rod-lens} and has no conical singularities.
\begin{itemize}
\item[(1)] $I_3= \{(\rho,z)\ | \ \rho=0,\ z>z_3\}$: \\
The Killing vector $v_3:=(0,0,1)=\partial/\partial\psi'$ vanishes. 
The condition for the absence of conical singularities on $I_3$ is given by 
\begin{eqnarray}
\lim_{\rho\to 0}\sqrt{\frac{\rho^2 f_1}{g_{\psi' \psi '} }}=\frac{\Delta \psi'}{2\pi}
\Longleftrightarrow 
C_f\frac{z_{30}-2b^2 }{z_{30}}=\left(\frac{\Delta \psi'}{2\pi}\right)^2  
\end{eqnarray}
for $z\in(z_3,\infty)$.
Hence, if we choose the periodicity of $\psi'$ as $\Delta \psi'=2\pi$, 
the condition can be satisfied on $I_3$ by setting 
\begin{eqnarray}
C_f=\frac{z_{30}}{z_{30}-2b^2 }.\label{eq:cf}
\end{eqnarray}

\item[(2)] $I_{0}= \{(\rho,z)\ | \ z_0<z<z_1,\rho=0\}$: \\
The Killing vector $v_{01}:=(0,1,0)=\partial/\partial\phi'$ vanishes. 
The condition for the absence of conical singularities on $I_0$ is given by
\begin{eqnarray}
\lim_{\rho\to 0}\sqrt{\frac{\rho^2 f_1}{g_{\phi'\phi'}} }=\frac{\Delta \phi'}{2\pi}
\Longleftrightarrow 
\frac{2C_fz_{10}^2(z_{30}-2b^2)}{z_{20}^2}=\left(\frac{\Delta \phi'}{2\pi}\right)^2. \label{eq:conical_01}
\end{eqnarray}

\item[(3)] $I_{2}= \{(\rho,z)\ | \ z_2<z<z_3,\rho=0\}$: \\
The Killing vector $v_{23}:=(0,1,\frac{bz_{21}}{z_{30}z_{32}-2b^2 z_{31}})=\partial/\partial\tilde \phi'$ vanishes. 
The conical-free condition is given by
\begin{eqnarray}
\lim_{\rho\to 0}\sqrt{\frac{\rho^2 f_1}{g_{1ab} v_{23}^av_{23}^b} }=\frac{\Delta \tilde\phi'}{2\pi}
\Longleftrightarrow 
\frac{2C_f(z_{30}z_{32}-2b^2z_{31})^2}{ z_{32} z_{31} (z_{30}-2b^2) }=\left(\frac{\Delta \tilde \phi'}{2\pi}\right)^2. \label{eq:conical_23}
\end{eqnarray}

\end{itemize}
Since $\phi'$ and $\tilde \phi'$ have the scale of length, it is useful to introduce angular coordinates $\varphi$ and $ \tilde \varphi$ with $2\pi$ periodicity by $\varphi:=L \phi'$ and $\tilde \varphi:=L \tilde \phi'$. Then, together with Eq.~(\ref{eq:cf}), the conditions (\ref{eq:conical_01}) and  (\ref{eq:conical_23}) can be written as, 
\begin{eqnarray}
 (\ref{eq:conical_01}) &\Longleftrightarrow&  \frac{2z_{10}^2z_{30}}{z_{20}^2}=L^2\left(\frac{\Delta \varphi}{2\pi}\right)^2 \\
  (\ref{eq:conical_23}) &\Longleftrightarrow&  \frac{2z_{30}(z_{30}z_{32}-2b^2z_{31})^2}{ z_{32} z_{31} (z_{30}-2b^2)^2 }=L^2\left(\frac{\Delta \tilde \varphi}{2\pi}\right)^2.
  \end{eqnarray}
  
  \medskip
Moreover, we impose the boundary  condition on the parameters $(z_i,b)$ so that the spatial topology of the horizon is lens space $L(n;1)=S^3/{\mathbb Z}_n$ ($n\in {\mathbb N}$).
From the mathematical discussion in Ref.~\cite{Hollands:2007aj}, this condition is represented by
\begin{eqnarray}
{\rm det}( \bar v_{01},\bar v_{23})=n
\Longleftrightarrow 
\frac{L b z_{21}}{z_{30}z_{32}-2b^2z_{31}}=n, 
\end{eqnarray}
where $( \bar v_{01},\bar v_{23})=L(  v_{01}, v_{23})$.

\medskip
(4) $I_{-}= \{(\rho,z)\ | \ z<z_0,\rho=0\}$: 
The Killing vector $v_{-}:=(1,0,0)=\partial/\partial t$ vanishes. 
This semi-infinite rod corresponds to an accelerating horizon.

\medskip
(5) $I_{1}= \{(\rho,z)\ | \ z_1<z<z_2,\rho=0\}$: 
The Killing vector $v_{1}:=(1,0,0)=\partial/\partial t$ vanishes. 
This finite rod corresponds to an event horizon.

\section{CTCs}\label{sec:ctcs}

We require absence of CTCs on $\Sigma \cup \partial \Sigma$.
The necessary and sufficient conditions to ensure that CTCs do not exist on $\Sigma \cup \partial \Sigma$ is such that $g_{\phi\phi}$ and $g_{\psi\psi}$ (or $g_{\phi'\phi'}$ and $g_{\psi'\psi'}$) become nonnegative in the region:
\medskip
The condition for the absence of CTCs is given by
\begin{align}
 \mu_1 W_{03}^2 W_{23}^2-4 b^2 \rho^2 \mu_0 \mu_2 W_{13}^2 > 0,
 \label{eq:noCTC}
\end{align}
which imposes an upper bound for $b^2$ at each point
\begin{align}
b^2 < U(\rho,z) :=\frac{\mu_1 W_{03}^2 W_{23}^2}{4 \rho^2  \mu_0 \mu_2 W_{13}^2}.
\end{align}

Therefore, if  the minimum $U_{min}$ of $U(\rho,z)$ exists on  $\Sigma \cup \partial \Sigma$  and $b^2< U_{min}$ holds,  CTCs do not exist in the  region. 
To prove this, we show that the function $U(\rho,z)$ has a minimum at $(\rho,z)=(0,z_3)$ on the rod  $I_\pm,I_i\ (i=0,\ldots,2)$. 
\medskip

It is not difficult to show that the function $U(\rho,z)$  has a minimum not on $\Sigma$ but on $\partial \Sigma$ . 
To see this, one should note that the norm of the gradient can be written as
\begin{align}
&( \partial_\rho U )^2 + (\partial_z U)^2= \frac{\mu_1^2 W_{03}^4 W_{23}^4}{4\rho^6 \mu_0^2 \mu_2^2 W_{00} W_{11} W_{22} W_{13}^4} \nonumber\\
& \qquad \times \left((\mu_0-\mu_1+\mu_2)^2 \rho^4 + 2 \mu_0\mu_2(2\mu_0 \mu_2-(\mu_0+\mu_2)\mu_1 + \mu_1^2)\rho^2+\mu_0^2\mu_1^2\mu_2^2 \right),
\end{align}
where  the first line is always positive for $\rho>0$, and the second line is also positive since 
\begin{align}
& (\mu_0-\mu_1+\mu_2)^2 \rho^4 + 2 \mu_0\mu_2(\mu_1^2-(\mu_0+\mu_2)\mu_1 + 2\mu_0 \mu_2)\rho^2+\mu_0^2\mu_1^2\mu_2^2\nonumber\\
& = \left(\left(\mu _0-\mu _1+\mu _2\right) \rho ^2+\frac{\mu _0  \mu _2 \left(\mu _1^2-(\mu_0+\mu_2)\mu_1+2 \mu_0\mu _2\right)}{\mu _0-\mu _1+\mu _2}\right)^2+\frac{4 \mu _0^3\mu_2^3 \left(\mu _0-\mu _1\right) \left(\mu _1-\mu _2\right) }{\left(\mu _0-\mu _1+\mu _2\right){}^2}, 
\end{align}
where the positivity of the last term can be shown from
\begin{align}
&(\mu_0-\mu_1)(\mu_1-\mu_2) 
\nonumber\\
&= 
\frac{z_{10} z_{21}(\mu_1 + \mu_0)(\mu_2 + \mu_1)}{(\sqrt{\rho^2+(z-z_1)^2}+\sqrt{\rho^2+(z-z_0)^2})(\sqrt{\rho^2+(z-z_2)^2}+\sqrt{\rho^2+(z-z_1)^2})}>0.
\end{align}
Therefore, the gradient of a smooth function $U(\rho,z)$  cannot be zero on $\Sigma$, which means that $U(\rho,z)$  
must have a minimum not on $\Sigma$ but on $\partial\Sigma$. 
Hence, in what follows, let us consider a minimum of $U(\rho,z)$ on  $\partial\Sigma$, which corresponds to $ I_i\ (i=\pm,0,\ldots,2), I_\infty$.

\medskip

First, let us consider the minimum of $U(0,z)$ on the rod $\rho=0$, i.e., on $I_i\ (i=\pm,0,\ldots,2)$. 
On $I_+$, we have
\begin{eqnarray}
U(0,z)=\frac{(z-z_0)(z-z_2)}{2(z-z_1)}, \quad
U_{,z}(0,z)=\frac{(z-z_1)^2+z_{10}z_{21}}{2(z-z_1)^2}>0,
\end{eqnarray}
and hence the monotonically increasing function $U(0,z)$ on $I_+$ has a minimum at $z=z_3$, which is written as
\begin{align}
U(0,z_3) = \frac{z_{30}z_{32}}{2  z_{31}}. \label{eq:localmin-1}
\end{align}
On the other hand,  since on $I_2$ 
\begin{eqnarray}
U(0,z)=\frac{(z-z_1)z_{30}^2z_{32}^2}{2(z-z_0)(z-z_2)z_{31}^2}, \quad 
U_{,z}(0,z)=-\frac{[(z-z_1)^2+z_{10}z_{21}]z_{30}^2z_{32}^2}{2(z-z_0)^2(z-z_2)^2z_{31}^2}<0,
\end{eqnarray}
the monotonically decreasing function $U(0,z)$ on $I_2$ has a minimum at $z=z_3$, and hence 
\begin{eqnarray}
U(0,z)\ge U (0,z_3).
\end{eqnarray}
Moreover, observing on $I_0$,  
\begin{eqnarray}
U(0,z)&=&\frac{(z-z_2)z_{30}^2}{2(z-z_0)(z-z_1)},\\
U_{,z}(0,z)&=&-\frac{[(z-z_2)^2-z_{20}z_{21}]z_{30}^2}{2(z-z_0)^2(z-z_1)^2}
\begin{cases}
<0 & ( z_0 <z< z_2-\sqrt{z_{20}z_{21}}) \\
>0  & (z_2-\sqrt{z_{20}z_{21}}<z<z_1),
\end{cases}
\end{eqnarray}
we find that the function $U(0,z)$ on $I_0$ has a local minimum at $z=z_*:=z_2-\sqrt{z_{20}z_{21}}$, and hence 
\begin{eqnarray}
U(0,z)\ge U (0,z_*)= \frac{z_{30}^2}{2(z_{20}+z_{21}-2 \sqrt{z_{20}z_{21}})},\label{eq:localmin-2}
\end{eqnarray}
where we note that the ratio of these minima on $I_0,I_2,I_+$ is computed as
\begin{align}
\frac{U(0,z_*)}{U(0,z_3)} = 
 \left(1+\frac{z_{21}}{z_{32}}\right) \left(1+\frac{z_{32}}{z_{20}}\right) \left(1-\sqrt{\frac{z_{21}}{z_{20}}}\right)^{-2}> 1.
\end{align}
Furthermore, near $I_-$ and $I_1$, the function $U(\rho,z)$ behaves as, respectively,  
\begin{eqnarray}
U(\rho,z)\simeq \frac{2(z_0-z)(z_2-z)(z_3-z)^2}{(z_1-z)\rho^2},\quad
U(\rho,z)\simeq \frac{2(z-z_1)(z_2-z)(z_3-z)^2z_{30}^2}{(z-z_0)z_{31}^2\rho^2},
\end{eqnarray}
which implies $U(0,z)=\infty$ on $I_-$ and $I_1$. 
To summarize, we have shown that the minimum of $U(0,z)$ on $I_i\ (i=\pm,0,\ldots,2)$ is given by Eq.~(\ref{eq:localmin-1}).

\medskip

 Next, let us consider the function $U(\rho,z)$ in the asymptotic region, namely, on  $I_\infty$.
We can show that in the asymptotic region, $U(\rho,z)$ behaves as
\begin{align}
 U(\rho,z) \simeq \frac{1}{1+\frac{z}{\sqrt{\rho^2+z^2}}}\sqrt{\rho^2+z^2}.
\end{align}
Hence, the function $U(\rho,z)$ diverges on $I_\infty$, so it cannot have a minimum on $I_\infty$.

Thus, we conclude that Eq.~(\ref{eq:localmin-1}) is also a minimum on $\Sigma\cup \partial \Sigma$ as well as on $\partial \Sigma$, and hence
the necessary and sufficient condition for the absence of CTCs is given by
\begin{align}
 b^2 < \frac{z_{30}z_{32}}{2z_{31}}. \label{eq:noctc-0}
\end{align}

\section{On existence of solutions} \label{sec:black lens}
 From the discussion in Sec~\ref{sec:boundary} and \ref{sec:ctcs}, we have shown that the absence of conical/orbifold  singularities, the black lens condition  require
\begin{align}
&\frac{2 z_{10}^2 z_{30}}{z_{20}^2} = L^2, \label{eq:regularity-con1}\\
& \frac{2 z_{30}( z_{32} z_{30}-2 b^2  z_{31})^2}{z_{32} z_{31}(z_{30}-2b^2)^2}=L^2, \label{eq:regularity-con2}\\
& \frac{L b z_{21}}{z_{30}z_{32}-2b^2 z_{31}} = n, \label{eq:regularity-con3}
\end{align}
Now, to confirm whether there is really a parameter range such that all these conditions can be satisfied, we reparameterize the rod interval $z_{i ,i-1}:=z_i-z_{i-1}\ (i=1,2,3)$ and the redefined BZ parameter $b$ as follows 
\begin{align}
z_{10} = \ell ,\quad  z_{21} := x \, \ell, \quad z_{32} = y \, \ell ,\quad L = \sqrt{\ell}\, \hat{L},\quad b = \sqrt{\ell} \, \hat{b},
\label{eq:regularity-param}
\end{align}
where $\ell$ fixes the size of the bubble on $I_0$, 
$x$ and $y$ are  the size of the horizon and the distance between the horizon and the center which is often called nut.
All the dimensionless parameters but $n$, $x,y,\ell,\hat L, \hat b$,  are assumed to be positive.
The condition for avoiding CTCs~\label{noctc-0} is now given by
\begin{align}
 b^2-\frac{y(1+x+y)}{2(x+y)}<0. \label{eq:noctc}
\end{align}
From Eq.~(\ref{eq:regularity-con3}), we have
\begin{align}
  \hat{L} = \frac{n ( y^2-2 \tilde{b}^2 (x+y) + y (1+x))}{\hat{b}\, x} \label{eq:regularity-ell}.
\end{align}
Eliminating $\hat{L}$ from Eqs.~(\ref{eq:regularity-con1}) and (\ref{eq:regularity-con2}) in terms of  Eq.(\ref{eq:regularity-ell}), we obtain
\begin{align}
&0= 2 \hat{b}^2 (x+y+1) \left\{ 2 n^2 (x+1)^2 y^2+2 n^2 (x+1)^2 x y+x^2\right \}\nonumber\\
&\qquad-4 \hat{b}^4 n^2 (x+1)^2
   (x+y)^2-n^2 (x+1)^2 y^2 (x+y+1)^2, \label{eq:regularity-con2a}
\end{align}
and
\begin{align}
&0=\left\{ y (x+y+1)-2 \hat{b}^2 (x+y)\right\}^2\nonumber\\
& \times \left[ 4 \hat{b}^4 n^2 y (x+y)-2 \hat{b}^2 (x+y+1) \left\{ 2 n^2y( x+y)+x^2\right\}+n^2 y (x+y) 
   (x+y+1)^2 \right]. \label{eq:regularity-con2b}
\end{align}

First, we consider the conical singularity free condition on $I_2$ in Eq.~(\ref{eq:regularity-con2b}),  which admits three branches for $\hat b^2$
\begin{eqnarray}
\hat b^2&=&\hat b_\pm^2:=\frac{(x+y+1) \left(x^2+2 n^2 (x+y)y \pm x \sqrt{x^2+ 4 n^2 x y+4 n^2 y^2}\right)}{4 n^2 (x+y)y}, \label{eq:b_pm}\\
\hat b^2&=&\hat b_0^2:=\frac{y(x+y+1)}{2(x+y)}.\label{eq:b_0}
\end{eqnarray}
From Eq.~(\ref{eq:b_pm}), we can show
\begin{eqnarray}
\hat b_\pm^2-\frac{y(1+x+y)}{2(x+y)}=\frac{x(x+y+1) \left(x+2n^2 y \pm  \sqrt{x^2+ 4 n^2 x y+4 n^2 y^2}\right)}{4 n^2 (x+y)y}>0,
\end{eqnarray}
where we note
\begin{eqnarray}
(x+2n^2 y)^2-(x^2+ 4 n^2 x y+4 n^2 y^2)=4n^2(n^2-1)y^2>0.
\end{eqnarray}
Therefore, this shows that the nonexistence condition of CTCs, (\ref{eq:noctc}),  cannot be satisfied for any $x>0$ and $y>0$. 
On the other hand, substituting  Eq.~(\ref{eq:b_0}) into Eq.~(\ref{eq:regularity-ell}), we can show
\begin{eqnarray}
\hat L=0,
\end{eqnarray}
which cannot satisfy $\hat L>0$. 
Hence, the solution without the conical singularity on $I_2$ cannot avoid CTCs.

\medskip
Next, we consider the conical singularity free condition on $I_0$ in Eq.~(\ref{eq:regularity-con2a}), from which we can obtain two branches for $\hat b^2$,
\begin{eqnarray}
\hat b^2=\tilde b^2_\pm:=\frac{(x+y+1)\left\{x^2+2n^2(x+1)^2(x+y) y \pm x\sqrt{x^2+4n^2(x+1)^2(x+y)y} \right\}}{4n^2(x+1)^2(x+y)^2},
\end{eqnarray}
which leads to
\begin{eqnarray}
\tilde b_\pm^2-\frac{y(1+x+y)}{2(x+y)}=\frac{x(x+y+1)\left(x\pm \sqrt{x^2+4n^2(x+1)^2(x+y)y} \right)}{4n^2(x+1)^2(x+y)^2}.
\end{eqnarray}
From these, we find that only the  branch  $\tilde b_-^2$ can satisfy the nonexistence condition of CTCs~(\ref{eq:noctc}).

\medskip
In summary, if one imposes the absence of conical singularities on the whole symmetry of axis $I_0,I_2,I_+$,  the presence of CTCs cannot be avoided around the center $(\rho,z)=(0,z_3)$. 
However, if one imposes  it only on $I_0$ and \blue{$I_+$}, one can obtain the solutions without CTCs, in which the horizon admits the lens space topology $L(n;1)$ for $n\ge 1$.

\section{Consistency with Chen-Teo static black lens}\label{sec:limit}
Here, we confirm that our solution coincides with the asymptotically flat, static black lens solution by Chen-Teo~\cite{Chen:2008fa} in  a certain  scaling limit, for which the following variables are used,
\begin{align}
 z_0 = - \lambda^2 ,\quad z_1 = - \lambda c \kappa^2,\quad z_2 = \lambda c \kappa^2,\quad z_3 = \lambda \kappa^2,\quad L = \sqrt{2} \lambda \bar{L},\quad a =  \frac{\bar{a}}{\sqrt{2}\kappa^2}.
\end{align}
With the rescaled coordinates,
\begin{align}
 \rho \to \lambda \bar{\rho},\quad z \to \lambda \bar{z},
\end{align}
and the rescaled parameters,
\begin{align}
\bar{z}_0 :=  z_1/\lambda = - c\kappa^2,\quad \bar{z}_1 := z_2/\lambda = c\kappa^2, \quad\bar{z}_1 :=z_3/\lambda = \kappa^2
\end{align}
the limit $\lambda \to \infty$ pushes $z_0$ away to $-\infty$, and one can see that the rod structure in $(\bar{\rho},\bar{z})$ recovers that of the static black lens in Ref.~\cite{Chen:2008fa}.
In the limit $\lambda \to \infty$, Eq.~(\ref{eq:regularity-con1})  requires
 \begin{align}
\bar{L} =1.
\end{align}
and then, Eqs.~(\ref{eq:regularity-con2}) and (\ref{eq:regularity-con3})  leads, respectively, to the nonexistence condition of the conical singularities  on $\bar z\in (\bar z_2,\bar z_3)$  and the condition of horizon topology $L(n;1)$  in Ref.~\cite{Chen:2008fa},
\begin{align}
\frac{(1-c-\bar{a}^2(1+c))^2}{(1-\bar{a}^2)^2(1-c^2)}=1,\quad 
\frac{2 \bar{a} c}{1-c-\bar{a}^2(1+c)}=n,
\end{align}
where $\bar{a}$ corresponds to $a$ in Ref.~\cite{Chen:2008fa}.
Moreover, the limit of the nonexistence condition for CTCs~(\ref{eq:noctc-0}) can be written as
\begin{align}
 \bar{a}^2 < \frac{c-1}{c+1},
\end{align}
which corresponds to the parameter region,  ``Region I",   in Ref.~\cite{Chen:2008fa}.  
Here, it should be noted that the solution in ``Region II" in \cite{Chen:2008fa}, which admits naked singularities and CTCs, 
are excluded from our solution by the nonexistence condition  (\ref{eq:noctc-0}) of CTCs.



\section{Summary and Discussions}\label{sec:summary}

In this paper, using the ISM for static and bi-axisymmetric Einstein equations, we have constructed the non-rotating black lens inside a bubble of nothing whose horizon is topologically lens space $L(n,1)=S^3/{\mathbb Z}_n$. 
Our work is entirely the parallel to the work of Chen-Teo~ \cite{Chen:2008fa}, where the static black ring as a seed solution is replaced with a static black ring in bubble of nothing~\cite{Astorino:2022fge}.
Using this solution, we have investigated whether a static black lens can be in equilibrium by the force balance between the expansion and gravitational attraction. 
If we require the absence of CTCs in the domain of outer communication, 
the non-rotating black lens must have conical singularities between the horizon and the center.  
It has been shown, however,  that for black lens, the existence of  expanding bubble does not exclude conical singularities and hence  the two forces,  the force of the bubble expansion and gravitational attraction, cannot be in static equilibrium unlike for the black ring.

\medskip
Refs.~\cite{Chen:2008fa,Tomizawa:2019acu}  have argued whether a rotating black lens can be in equilibrium by the balance between 
the gravitational force (attraction) and the centrifugal force (repulsive force), and concluded that it cannot be in equilibrium. 
The generalization of this rotating black lens solution to one in expanding bubble may be an interesting issue, 
since whether such a  black lens without conical singularities exists depends on the balance among the gravitational force, the centrifugal force and the expansion of the bubble. 
Such a rotating black lens solution is expected to be two or more than soliton solution. 
This deserves our future work.




\acknowledgments

This work is supported by Toyota Technological Institute Fund for Research Promotion A. 
RS was supported by JSPS KAKENHI Grant Number JP18K13541. 
ST was supported by JSPS KAKENHI Grant Number 21K03560.




\end{document}